\documentclass[11pt]{article}

\usepackage{latexsym}
\usepackage{amsfonts}
\usepackage{amssymb}
\usepackage{amsmath}
\usepackage{amsthm}

\newcommand{\bea}{\begin{eqnarray}}
\newcommand{\eea}{\end{eqnarray}}
\newcommand{\beann}{\begin{eqnarray*}}
\newcommand{\eeann}{\end{eqnarray*}}
\newcommand{\beq}{\begin{equation}}
\newcommand{\eeq}{\end{equation}}
\newcommand{\ba}{\begin{array}}
\newcommand{\ea}{\end{array}}
\newcommand{\ben}{\begin{enumerate}}
\newcommand{\een}{\end{enumerate}}
\newcommand{\bit}{\begin{itemize}}
\newcommand{\eit}{\end{itemize}}

\numberwithin{equation}{section}

\def\half{{\frac{1}{2}}}

\newcommand{\asy}{\longrightarrow}
\newtheorem{theorem}{Theorem}

\def\cH{{\cal H}}

\begin{document}

\title{
\vspace*{-2cm}
\begin{flushright}
\normalsize{
ULB-TH/03-26\\
MPI-MIS-51/2003\\
gr-qc/0306112 }
\end{flushright}
\vspace{1.5cm} Asymptotically anti-de Sitter space-times:
symmetries and conservation laws revisited \vspace*{.5cm}}

\author{Glenn Barnich$^{a}$\thanks{Research
Associate of the National Fund for Scientific Research
(Belgium).}, Friedemann Brandt$^b$ and Kim Claes$^{a}$
\vspace{.3cm}
\\
$^a$Physique Th\'eorique et Math\'ematique, Universit\'e Libre de
Bruxelles,\\
Campus Plaine C.P. 231, B--1050 Bruxelles, Belgium \vspace{.3cm}
\vspace{.3cm}\\
$^b$Max-Planck-Institute for Mathematics in the Sciences,\\
Inselstra\ss e 22--26, D--04103 Leipzig, Germany \vspace{.3cm}}
\date{}
\maketitle

\begin{abstract}
In this short note, we verify explicitly in static coordinates
that the non trivial asymptotic Killing vectors at spatial
infinity for anti-de Sitter space-times correspond one to one to
the conformal Killing vectors of the conformally flat metric
induced on the boundary. The fall-off conditions for the metric
perturbations that guarantee finiteness of the associated
conserved charges are derived. \vspace{1pc}
\end{abstract}

\section{Exact Killing vectors and their fall-off}

\nocite{HawkingEllis,Abbott:1982ff,Henneaux:1985tv,Brown:1986nw,%
Henneaux:1985ey,Anderson:1996sc,Aharony:1999ti,Witten:1998qj,Ashtekar:1999jx}

Consider the $(n+1)$-dimensional flat embedding space with
coordinates $X^\Delta$, $\Delta=0,\dots,n$ and metric
$\eta_{\Delta\Gamma}=\mathrm{diag}(-1,-1,1,\dots,1)$ for $n\geq 3$.
Anti-de Sitter
space-time can be defined as the hypersurface
\bea 
\eta_{\Delta\Gamma}X^\Delta X^\Gamma=-l^2.\eea
Static coordinates $x^\mu$, $\mu=0,\dots,n-1$ on (universal)
anti-de Sitter space-time can be chosen as
$x^\mu\equiv
\tau,r,
y^A$, $A=2,\dots, n-1$, with \bea\begin{array}{rcl}
X^0&=&l(1+\frac{r^2}{l^2})^{\half}
{\sin(\,{\tau}/{l})},\\
X^1&=&l(1+\frac{r^2}{l^2})^{\half}\cos({\tau}/{l}),\\
X^2&=&r\cos y^2,\\
X^3&=&r\sin y^2\cos y^3,\\
&\vdots&\\
X^{n-1}&=&r\sin y^2\dots\sin y^{n-2}\cos y^{n-1},\\ X^{n}&=&r\sin
y^2\dots\sin y^{n-1}.
\end{array}\eea
In these coordinates, the metric on anti-de Sitter
space-time is given by \bea d\bar s^2\equiv \bar g_{\mu\nu}dx^\mu
dx^\nu=-(1+\frac{r^2 }{l^2})d\tau^2+\nonumber\\+(1+\frac{r^2
}{l^2})^{-1}dr^2
+r^2\sum_{A}f_A(dy^A)^2, \eea with $ f_2=1$,
$f_A=\sin^2 y^2\sin^2 y^3\dots\sin^2 y^{A-1}$ for $3\leq A\leq
n-1$.

The Killing vectors for this metric can be obtained as follows:
using the metric $\eta_{\Delta\Gamma}$ to lower the indices, the
Killing vectors of the flat embedding space are
$\xi_\Delta=a_\Delta+b_{\Delta\Gamma} X^\Gamma$, with constant
$a_\Delta,b_{\Delta\Gamma}=-b_{\Gamma\Delta}$. Only the vectors
corresponding to the Lorentz transformations are tangent to the
hypersurface. The Killing vectors of $d\bar s^2$ are then given by
\bea \xi^\alpha=\bar g^{\alpha\beta} \xi_\beta,\
\xi_\beta=b_{\Delta\Gamma}\frac{\partial X^\Delta}{\partial
x^\beta}X^\Gamma,\eea or explicitly,
\bea\xi^\tau=-b_{01}l-b_{0A}X^A(1+\frac{r^2
}{l^2})^{-\half}\cos({\tau}/{l}) +b_{1A}X^A(1+\frac{r^2
}{l^2})^{-\half}{\sin({\tau}/{l})},\eea \bea \xi^r=-l(1+\frac{r^2
}{l^2})^{\half}\Big[{\sin({\tau}/{l})} (b_{0A}\frac{X^A}{r})+
\cos({\tau}/{l})(b_{1A}\frac{X^A}{r})\Big],\eea \bea \xi^A=
\frac{1}{r^2f_A}\frac{\partial X^B}{\partial y^A}\Big[b_{B0}l
(1+\frac{r^2 }{l^2})^{\half}{\sin({\tau}/{l})} + b_{B1}l
(1+\frac{r^2 }{l^2})^{\half}\cos({\tau}/{l}) +b_{BC}X^C\Big]. \eea
Let $a=0,2,\dots, n-1$, with $x^0=\tau, x^A=y^A$. At spatial
infinity $r\asy +\infty$, the asymptotic fall-off conditions for
these Killing vectors are \bea\xi^r\asy O(r),\ \xi^a\asy
O(r^0).\label{fokv}\eea

\section{Asymptotic Killing vectors and conformal Killing vectors on the boundary}

\subsection{Asymptotic Killing vectors of anti-de Sitter space-times}

For arbitrary vector fields
$\xi^\mu$ that satisfy the fall-off conditions (\ref{fokv}) of the exact Killing vectors,
as well as $\xi^\mu\asy O(r^m)\ \Rightarrow\
\partial_r\xi^\mu\asy O(r^{m-1})$,
we determine the fall-off of $L_\xi \bar g_{\alpha\beta}$.
Explicitly, \bea L_\xi \bar g_{rr}\asy O(r^{-2}),\, L_\xi\bar
g_{ar}\asy O(r),\, L_\xi \bar g_{ab}\asy O(r^{2}).\eea This
motivates us to define asymptotic Killing vectors through the
fall-off conditions (\ref{fokv}) and the constraints \bea &L_\xi
\bar g_{rr}\asy o(r^{-2}),\, L_\xi\bar g_{ar}\asy o(r),\, L_\xi
\bar g_{ab}\asy o(r^{2}). \label{asykilling}\eea We remark that
this definition does not use assumptions on fall-off conditions
for the metric perturbations. In fact, eventually we shall derive
such conditions by requiring the existence of asymptotically
conserved $(n-2)$-forms associated to the asymptotic Killing
vectors. Regarding this point, we thus reverse the approach used
in \cite{Barnich:2001jy} where we derived fall-off conditions
determining asymptotic Killing vectors from fall-off conditions
for the fields.

Asymptotic Killing
vectors that fall-off as \bea \xi^r\asy o(r),\ \xi^a\asy o(r^0)
\eea automatically satisfy (\ref{asykilling}) and are considered trivial.
We want to compute the equivalence classes of
asymptotic Killing vectors modulo trivial ones.

Non trivial Killing vectors can thus be parameterized by \bea
\xi^\tau\asy lT(\tau,y),\ \xi^r\asy rR(\tau,y)+o(r),\  \xi^A\asy
\Phi^A(\tau,y).\label{para}\eea The constraints (\ref{asykilling})
then reduce to \footnote{An index in parentheses indicates that
the summation convention does not apply.} \bea
&R=-lT_{,\tau},\label{a0}\\ &T_{,A}=lf_{(A)}\Phi^A_{,\tau}\label{a3},\\
&lT_{,\tau}=\Phi^A_{,(A)}+\sum_{B<A}\Phi^B\cot y^B,\label{a1}\\
&\Phi^B_{,A}f_{(B)}+\Phi^A_{,B}f_{(A)}=0,\ A\neq B\label{a4}. \eea

\subsection{Conformal Killing vectors on the boundary}

For large $r$, the metric $d\bar s^2$ behaves to leading order as
$d\bar s^2\asy \frac{r^2}{l^2}(d{s^\prime}^2)$ where \bea
d{s^\prime}^2= g'_{ab}dx^a
dx^b=-d\tau^2+l^2\sum_{A=2}^{n-1}f_A(dy^A)^2\eea and is thus
asymptotically conformal to this latter metric, which will be
referred to as the "metric induced on the boundary" below.

The conformal Killing vectors of the metric induced on the boundary
satisfy \bea L_\xi
g'_{ab}=\frac{2}{n-1}\,D'_c\xi^c\, g'_{ab}.\eea
Using the notation%
\footnote{At this stage it is actually not yet clear that
the functions $T(\tau,y)$ and $\Phi^A(\tau,y)$
coincide with those in (\ref{para}). Nevertheless
we employ the same notation.}
$\xi^\tau= lT(\tau,y)$ and $\xi^A= \Phi^A(\tau,y)$ one obtains
\bea
\lefteqn{
D'_c\xi^c=\partial_c \xi^c+\frac{1}{2}
\sum_{A=2}^{n-1}\xi^c\partial_c
{\ln{f_A}}
}
\nonumber\\
&&=lT_{,\tau}+\sum_{A=2}^{n-1}(\Phi^A_{,A}+\sum_{B<A}\Phi^B \cot
y^B).\eea Explicitly, this gives the conditions \bea
&&lT_{,\tau}=\frac{1}{n-1}\Big(lT_{,\tau}+\sum_{A=2}^{n-1}
(\Phi^A_{,A}+\sum_{B<A}\Phi^B\cot y^B)\Big)\label{c1}, \\
&&\Phi^A_{,(A)}+\sum_{B<A}\Phi^B\cot
y^B=\frac{1}{n-1}\Big(lT_{,\tau}+\sum_{A=2}^{n-1}
(\Phi^A_{,A}+\sum_{B<A}\Phi^B\cot y^B)\Big)\label{c2}, \\
&&T_{,A}=lf_{(A)}\Phi^A_{,\tau},
\label{c3}\\
&&\Phi^B_{,A}f_{(B)}+\Phi^A_{,B}f_{(A)}=0,\ A\neq B.
\label{c4} \eea

\subsection{Correspondence}

\begin{theorem}
The non trivial asymptotic Killing vectors are in one to one
correspondence with the conformal Killing vectors of the metric
$g'_{ab}$ induced on the boundary.
\end{theorem}
{\bf Proof:} The asymptotic Killing equations imply the conformal
ones for the boundary metric. Indeed, (\ref{c1}) is equivalent to
\bea lT_{,\tau}=\frac{1}{n-2}\sum_{A=2}^{n-1}
(\Phi^A_{,A}+\sum_{B<A}\Phi^B\cot y^B),\label{c1'}\eea which is
implied by the sum over $A$ of (\ref{a1}). Using (\ref{c1'}),
(\ref{c2}) is equivalent to \bea \Phi^A_{,(A)}+\sum_{B<A}\Phi^B
\cot
y^B=\frac{1}{n-2}\sum_{A=2}^{n-1}(\Phi^A_{,A}+\sum_{B<A}\Phi^B\cot
y^B),\label{c2'}\eea which is implied by using (\ref{a1}) in
(\ref{c1'}). Finally (\ref{c3}) is (\ref{a3}) and (\ref{c4}) is
(\ref{a4}).

Conversely, the conformal Killing vectors imply the asymptotic
ones. Indeed, we only need to verify that (\ref{a1}) hold
(since (\ref{a0}) only determines $R$ and does not
impose any constraint). This follows from using (\ref{c2'}) in
(\ref{c1'}).\qed

\subsection{Discussion}

The boundary metric is conformal to the flat
Minkowski metric $\eta_{ab}$, so that the conformal Killing
vectors of the boundary metric, and thus also the equivalence
classes of asymptotic Killing vectors, correspond one to one to the
conformal Killing vectors of Minkowski space in $n-1$ dimensions.
For $n=3$ one thus gets the infinite dimensional pseudo conformal
algebra in 2 dimensions. There is thus symmetry enhancement as the exact Killing
vectors are given by the 6 dimensional $so(2,2)$ algebra. For $n>3$, one gets an algebra that is
isomorphic to $so(n-1,2)$ of dimension $\frac{(n+1)n}{2}$, and thus
a one to one correspondence between exact and non trivial asymptotic
Killing vectors.

\section{Charges and fall-off conditions}

Let $g_{\mu\nu}=\bar g_{\mu\nu}+h_{\mu\nu}$, $h=g^{\mu\nu}h_{\mu\nu}$,
and $S^{\infty}_{n-2}$ the $(n-2)$-dimensional sphere at spatial infinity defined by
$\tau=\mathrm{constant}$, $r=R\rightarrow \infty$, and \beann(d^{n-p}x)_{\mu_1\dots\mu_p}=\frac{1}{p!(n-p)!}
\epsilon_{\mu_1\dots\mu_n}dx^{\mu_{p+1}}\dots dx^{\mu_n},\eeann with
$\epsilon_{0\dots n-1}=1$. We restrict the discussion to
functions $h_{\mu\nu}$ which depend on $r$ such that
$h_{\mu\nu}\asy O(r^m)$ implies $\partial_r h_{\mu\nu}\asy
O(r^{m-1})$ and $\partial^2_r h_{\mu\nu}\asy O(r^{m-2})$ for $r\asy\infty$.

We shall now discuss the charges associated with the
asymptotic Killing vectors, see, e.g.,
\cite{HawkingEllis,Abbott:1982ff,%
Henneaux:1985tv,Brown:1986nw,Henneaux:1985ey,Anderson:1996sc,%
Ashtekar:1999jx,Barnich:2001jy}. Our discussion is based on
results in the last of these references where the following
expression for these charges was derived: \bea
Q_{\xi}=-\int_{S^{\infty}_{n-2}}(d^{n-2}x)_{\nu\mu}k^{\nu\mu}_\xi
=-\lim_{r\asy\infty}\int\prod_{A=2}^{n-1}dy^Ak^{[\tau r]}_\xi
\label{Q} \eea with \bea k^{\nu\mu}_\xi[h;\5g] =-\frac{\sqrt{-\bar
g}}{16 \pi}\Big[\bar D^\nu (h\xi^\mu)+\bar
D_\sigma(h^{\mu\sigma}\xi^\nu) +\bar
D^\mu(h^{\nu\sigma}\xi_\sigma) +\frac{3}{2}h\bar
D^\mu\xi^\nu\nonumber\\+\frac{3}{2} h^{\sigma\mu}\bar
D^\nu\xi_\sigma+\frac{3}{2} h^{\nu\sigma}\bar D_\sigma\xi^\mu
-(\mu\leftrightarrow \nu)\Big]. \label{gsuperpot0}\eea As shown in
\cite{Barnich:2001jy}, in order that these charges are meaningful,
the $(n-2)$-forms $(d^{n-2}x)_{\nu\mu}k^{\nu\mu}_\xi$ must be
asymptotically conserved at $S^{\infty}_{n-2}$ which imposes \bea
d^nx\,\cH^{\mu\nu}\,L_\xi \bar g_{\mu\nu}\asy 0,
\label{ourcondition} \eea where $\cH^{\mu\nu}$ are the left hand
sides of the linearized field equations associated to the action
$S=\int d^nx \sqrt{-g}(R-2\Lambda)$, with
$\Lambda=-\frac{(n-1)(n-2)}{l^2}$. In our present approach,
(\ref{ourcondition}) determines the fall-off conditions for the
functions $h_{\mu\nu}$ to which the results apply. To derive these
conditions, we use that (\ref{asykilling}) and
(\ref{ourcondition}) impose \bea \cH^{rr}\asy O(r),\ \cH^{ar}\asy
O(r^{-2}),\ \cH^{ab}\asy O(r^{-3}). \label{Hfalloff} \eea Using
now the explicit expression for $\cH^{\mu\nu}$, \bea
\cH^{\mu\nu}[h;\5g]= \frac{\sqrt{-\bar g}}{32
\pi}\Big(\frac{2\Lambda}{n-2}(2h^{\mu\nu} -\bar g^{\mu\nu}h) +\bar
D^\mu\bar D^\nu h+\bar D^\lambda\bar D_\lambda h^{\mu\nu} -2\bar
D_\lambda \bar D^{(\mu}h^{\nu)\lambda} \nonumber\\-\bar
g^{\mu\nu}(\bar D^\lambda\bar D_\lambda h - \bar D_\lambda\bar
D_\rho h^{\rho\lambda})\Big), \label{Hmunu}\eea we can eventually
derive the fall-off behavior that generic functions $h_{\mu\nu}$
must have in order that (\ref{Hfalloff}) holds. The result is \bea
h_{rr}\asy O(r^{-n-1}),\, h_{ar}\asy O(r^{-n}),\, h_{ab}\asy
O(r^{-n+3}). \label{newlabel} \eea These conditions agree with
those imposed in
\cite{Henneaux:1985tv,Brown:1986nw,Henneaux:1985ey} and imply the
finiteness of the charges (\ref{Q}).

\section*{Acknowledgments}
The work of  GB is supported in part by the ``Actions de Recherche
Concert{\'e}es'' of the ``Direction de la Recherche
Scientifique-Communaut{\'e} Fran\c{c}aise de Belgique, by a
``P{\^o}le d'Attraction Interuniversitaire'' (Belgium), by
IISN-Belgium (convention 4.4505.86), by Proyectos FONDECYT 1970151
and 7960001 (Chile) and by the European Commission RTN program
HPRN-CT00131, in which the author is associated to K.~U.~Leuven.

\providecommand{\href}[2]{#2}\begingroup\raggedright\endgroup


\end{document}